\documentclass[preprint,prd,aps,superscriptaddress,tightenlines,nofootinbib,showpacs]{revtex4}
\usepackage{graphics,supertabular}
\input psfig.sty

\def\sqr#1#2{{\vcenter{\vbox{\hrule height.#2pt
 \hbox{\vrule width.#2pt height#1pt \kern#1pt
 \vrule width.#2pt} \hrule height.#2pt}}}}

\def\operp{\hbox{${\kern+.25em{\bigcirc}
\kern-.85em\bot\kern+.85em\kern-.25em}$}}

\def\lsim{\;\raise0.3ex\hbox{$<$\kern-0.75em\raise-1.1ex\hbox{$\sim$}}\;}
\def\gsim{\;\raise0.3ex\hbox{$>$\kern-0.75em\raise-1.1ex\hbox{$\sim$}}\;}

\def\ve{\vfill\eject}
\def\rdots{\mathinner{\mkern1mu\raise1pt\vbox{\kern7pt\hbox{.}}\mkern2mu
\raise4pt\hbox{.}\mkern2mu\raise7pt\hbox{.}\mkern1mu}}

\def\e e{$e^+ e^-$ }

\begin{document}
\large
\title{QCD INEQUALITIES AND THE $D_s(2320)$}
\author{S. Nussinov$^{1,2,3}$}

\affiliation{$^1$School of Physics and Astronomy, \\
         Tel Sackler Faculty of Exact Sciences,\\
         Tel Aviv University and \\
         $^2$Dept. of Physics \&\ Astronomy,\\
         Univ. of California, Los Angeles, CA 90095\\
         $^3$Dept. of Physics \&\ Astronomy,\\
         Univ. of South Carolina, SC 29208}
\date{\today}

\begin{abstract}
      We discuss the new state $D_{sJ}(2317)$ discovered by BaBar and
demonstrate using QCD inequalities that if indeed the $f_0(980)$ and the
new $D_{sJ}(2317)$ ($0^+$) are primarilly made of four quarks that a new
I=0 ``$\bar DD$ bound state'' at a mass smaller than 3660 MeV must exist.
Observation of such a state will constitute definitive evidence for
four-quark states. 
\end{abstract}

\maketitle

\section {General Phenomenological Consideration.}
\vskip.3cm

The new $D_s(2320)$ discovered in BaBar\cite{BaBar} as a narrow $D_s(1970) + \pi^0$
resonance is somewhat puzzling\cite{BCL, CJ, BEH, CdF}.  It has $c\bar s$ flavors and
natural spin parity $J^P = 0^+,1^-,2^+\ldots$.  We assume $J^P=0^+$
as suggested by the lightness of the state.  The only allowed two-body decay
is then $D_s(1963) + \pi^0$. $D_s(1963) + \eta$ is $\sim 40$ MeV above
2320 MeV and $D_s(2320)\to D_s(1963) + \gamma$ is a forbidden
$0\to 0$ electromagnetic transition.

Despite the high $Q$ value in $D_s(2320)\to D_s(1968) + \pi$ 
the narrow width of the $S$-wave decay immediately follows if the
$I$-spin of $D_s(2320)$ is $I=0$.  The final $D_s+\pi^0$ state has $I=1$
and the decay rate would naturally be suppressed to be
$\Gamma(D_s(2320)) \leq$ expt. resolution $\approx$ 10 MeV.  We will
henceforth assume $I\{D_s(2320)\} = 0$ which is the case if
$D_s(2320)$ is just a $c\bar s$ state and/or $c\bar s$ with any number of
gluons or $\bar uu + \bar dd~, I=0$ light quark pairs.

The puzzling feature is the relative lightness of this state.  The simplest
assignment of $D_s(2320)$ to be the ${}^3P(0^+)$, $P$-wave quark model
$c\bar s$ state conflicts with earlier quark model
calculations.\cite{GI, GK} These
models treat the system as a $q\bar Q$ state and accommodate the
$^1S$(1963) $D_s$ and the 
$^3S(2112)~D_s^*$ states and a pair of $P$-wave states with
$j_s = \ell_s + s_s = 3/2$ identified with the narrow $D_{s_J}(2573)$ and
$D_{s_1}(2536)$ with likely $J^P = 2^+,1^+$ spin parities.  The remaining
two $P$-wave states corresponding to $j_s = 1/2$ with $j^P = 1^+0^+$ were
predicted to lie above 2400 MeV and hence the $0^+$ state was expected to
decay strongly into $DK$ and be very broad.  These considerations, on which
we elaborate further, make a $P$-wave $c\bar s$ interpretation of
$D_s(2320)$ difficult. Yet, such an interpretation is deemed to be
favored by the recent discovery of the $J^P = 1^+$ $D_s(2460)$.\cite{CLEO, Belle}

We note that for all known $Q\bar q$, $q \bar q$, and $Q\bar Q$ states
the gap between the ${}^3S$ state and the lowest $P$ state exceeds
$\approx$ 320 MeV, making the $\approx$ 220 MeV split between
$D_s^*(2112)$ and $D_s(2320)$ clearly stand out as the smallest.

This holds also for the new $0^+$ $c\bar u$ state seen in Belle\cite{Belle}
as a broad resonance in the $D\pi$ channel at 2308 MeV (again 300 MeV
above the triplet S-wave $D^*(2010)$.

If indeed the new $0^+$ $c\bar u$ state is confirmed the lightness
of the $D_{sJ}(2317)$ is even more puzzling as it is almost degenerate
with
its non-strange $c\bar u$ analog, whereas in all 
meson/baryon states the penalty for such a $u\to s$ substitution exceeds
$\sim$100-150 MeV, the constituent quark $s-u$ mass difference.

The large splitting are indeed suggested by the $P$-wave centrifugal
barrier: ${2\over\mu}\langle{1\over r^2}\rangle$ with $\mu =
m_q/2$ or $\sim m_q$, $m_Q/2$ for $\bar qQ$ or $\bar qQ~\bar QQ$
(since $\langle{1\over r^2}\rangle \geq {1\over<r^2>}$ increases with
$\mu$, this splitting changes only mildly with $\mu$).

The above suggests considering $D_s(2320)$ as
primarily an $I=0$ four-quark $c\bar s(\bar uu + \bar dd)/\sqrt{2}$ bound
state or the $I=0$ $(K^+D^0 + K^0D^+)/\sqrt{2}$ bound state.

The idea that certain states can be more readily obtained as four-quark
states and considerations of exotic
(non-$\bar qq$) or $qqq$) states in general date back to the very beginning
of quark models and QCD.  Baryonium $(qq)\ldots (\bar q \bar q)$ states
were suggested by considerations of duality \cite{Ros} and in recent years
re-appeared as light-heavy $(QQ^\prime)(\bar q\bar q^\prime)$
tetra quarks and/or as molecular bound states of $DD^*~DD^*$
or even $KK^*$ dominated by pion exchange \cite{T, MW, GN}.

Since the lightest $P$-wave $\bar ss$ state is expected to lie significantly
above the ${}^3S(\bar ss)~\phi(1020)$, the light $0^{++}I=0(980)$ and
$I=1~a_0(980)$ are unlikely to be just $\bar ss$ state. R. Jaffe
suggested \cite{Jaf77a} 
that these states arise as four-quark states within bag models, thanks to a favorable pattern
of chromomagnetic ``hyperfine" attraction.  This in turn motivated prediction
of striking, strong interactions stable, Hexa $u^2s^2d^2$ and Penta--$\bar csu^2d$
states--which were not found to date.\cite{Jaf77b, Lip, ESGR}

In general the paucity of exotic $\bar qq\bar qq$ and $q^3\bar qq$ mesons
and baryons traces to the lightness of the pseudoscalar pion 
causing the exotics to be extremely broad and overlapping resonances.  The
decay of non-exotic $\bar qq$ states say $\rho$ into a pair of $\bar qq$
pions is suppressed by a $1/N_c$ color factor.\cite{Wit79} This is {\bf not} the case
for decay of exotic hadrons which simply fall apart into its constituent
non-exotic hadrons with no penalty for creating an extra $\bar qq$ pair.

Thus if the new $D_s(2320)$ is not a standard $P$-wave $c\bar s$ state, it
may well be the first genuine exotic four quark $c\bar s{\bar uu+\bar dd\over
\sqrt{2}}$ or $(D^0K^+ + D^+K^0)/\sqrt{2}$ which is narrow. 

Our main aim in the following is to point out a likely analogous $c\bar c{\bar uu+\bar dd\over
\sqrt{2}}$ or $D\bar D$ state. QCD inequalities then suggest that this
new state is lighter than 3660 MeV.

More likely
$D_s$ is the appropriate lowest energy superposition of all three:

\begin{equation}
D_s(2320) = \alpha\bigl|c\bar s\rangle + \beta\bigl|c\bar s{(\bar uu+\bar dd)
\over\sqrt{2}}\rangle + \gamma\bigl|{K^+D^0+K^0D^+\over\sqrt{2}}\rangle
\label{e01}
\end{equation}

with $|\beta|^2 + |\gamma|^2 > |\alpha|^2$ favored by our interpretation.

\vskip.5cm

\section {QCD/Lattice Formulation of the Hadron Spectrum.} 
\vskip.3cm

The above short summary fails to address the main issue.  A reliable
theoretical framework for computing the hadronic spectrum in general
(and of four-quark states in particular) is lacking.

In principle, one could use a path integral formulation together with an
appropriate lattice discretization to numericaly evaluate Euclidean
$n$-point correlators of local color singlet operators.\cite{NL} 

To find the mass of the lightest hadron in a channel with given quark flavors
and Lorentz $(J^P)$ quantum numbers, we use the two-point correlator of
currents with these quantum numbers:

\begin{equation}
\langle 0|\{\bar\psi_i(x)\Gamma\psi_j(x)\}^+\{\bar\psi_i(y)\Gamma\psi_j(y)\}|0\rangle
\equiv F_{(\Gamma)}^{(ij)}(x,y)  
\label{e02}
\end{equation}

with $\Gamma = 1,\gamma_\mu,\gamma_\nu\gamma_5,\sigma_{\mu\nu}\gamma_5$
the approprite Dirac matrix.  The asymptotic behavior of $F$ as
$|x-y|\to\infty$ is controlled by the lowest mass particle--say the
$0^+(c\bar s)$ scalar state of interest when we use $\Gamma = 1~~
i=c,j=s$ :

\begin{equation}
F^{(c\bar s)}(x,y) \buildrel |x\to y|\to\infty \over \sim
|<0[\bar c(0)s(0)]D_s^{(0^+)}(2320)\rangle|^2\cdot
e^{-m^{(0)}_{c\bar s}|x-y|}
\label{e03}
\end{equation}

More generally hadron masses, couplings and scattering amplitudes can be
obtained by inverse Laplace transforms of two, and $n$-point Euclidean
correlators. The latter can be written after integration over the
fermionic fields (which appear in the action in a bilinear form) as
path integrals over gauge field configurations $A_\mu^a(x)$.

For $F_\Gamma^{(ij)}(x,y)$ we have:

\begin{equation}
F_\Gamma^{(ij)}(x,y) = \int d^\mu(A) {\rm tr}\{\Gamma S_i^A(x,y)
\Gamma S_j^A(y,x)\}
\label{e04}
\end{equation}

with $S_i^A(x,y)$ the propagator of the quark $q_i$, in the background
field $A_\mu^a(x)$, the trace is over spinor and color indices and
$d\mu(A)$ is the (positive!) measure obtained after integrating the fermions
to yield the determinant factor

\begin{equation}
d\mu(A) = D(A_\mu^{(a)}(x)) e^{-S_{YM}(A_\mu(x))}
\prod^{n_F}_{i=1}~{\rm det}(D\!\!\!/_A+m_i)
\label{e05}
\end{equation}

and $S_{YM} = \int d^4x{\cal{L}}_{YM} = \int d^4x(\vec E^2+\vec B^2)$
is the Euclidean Yang-Mills action.

Unfortunately, the computational complexity of generating many gauge
backgrounds and evaluating for each the fermionic propagator and, in
particular, the fermionic determinant prevents carrying out these calculations
reliably at present.  Note that since the state of interest is likely to contain extra $\bar qq$ pairs $\{(\bar uu + \bar dd)/\sqrt{2}\}$ the
``quenched" approximation in which ${\rm det}(D\!\!\!/_A+m_i)\to 1$ and
quark loops are omitted may be inappropriate.

To bring out such components in the $D_s(2320)$ state we can use instead of
the correlators of quark bilinears products of quark
bilinear such as:

\begin{equation}
F_{s\bar c,\bar uu}(x,y)\sim\langle 0|[\bar c(x)\gamma_5
s(x)\cdot(\bar u(x)\gamma_5u(x))]^+\cdot\bar c(y)\gamma_5s(y)\cdot
(\bar u(y)\gamma_5u(s))|0\rangle 
\label{e06a}
\end{equation}

or

\begin{equation}
F_{s\bar u,c\bar s}(x,y)\sim\langle 0|[\bar c(x)\gamma_5
u(x)\cdot(\bar u(x)\gamma_5s(x))]^+\cdot{[\bar c(y)\gamma_5c(y))
(\bar u(y)\gamma_5(y)]}|0\rangle 
\label{e06b}
\end{equation}

After integrating the fermionic fields and using $\gamma_5S^A(x,y)\gamma_5 =
(S^A(y,x))^+$ with the + indicating the adjoint color/spin matrix (\ref{e06a}) and
(\ref{e06b}) yields the following path integrals:

\begin{eqnarray}
&\omit&F_{s \bar c\bar uu}(x,y) = \int d\mu(A)~{\rm tr}((S_c^A(x,y)^+
S_s^A(x,y))~{\rm tr}((S_u^A(x,y)^+S_u^A(x,y)) \nonumber \\
                        &+& \int d\mu(A)~{\rm tr}(S_c^A(x,y)^+S_s^A(x,y))~{\rm tr}
(S_u^A(x,x)\gamma_5)\cdot{\rm tr}(S_u^A(y,y)\gamma_5)
\label{e07a}
\end{eqnarray}

or

\begin{eqnarray}
&\omit&F_{s \bar uc\bar s}(s,y) = \int d\mu(A)~{\rm tr}(S_c^A(x,y)^+
S_u^A(x,y)~{\rm tr}(S_u^A(x,y)^+S_s^A(x,y)) \nonumber \\
      &+& \int d\mu(A)~{\rm tr}(S_c^A(x,y)^+S_s^A(x,y))~{\rm tr}(S_u^A(x,x)\gamma_5))
\cdot{\rm tr}(S_u^A(y,y)\gamma_5) 
\label{e07b}
\end{eqnarray}

Note that (\ref{e06a}) and (\ref{e07a}) suggest viewing $D_s(2320)$ as a ``$D_s\eta$" composite
whereas (\ref{e06b}) and (\ref{e07b}) emphasize the more relevant ``$DK$" component.
Also instead of $\bar c(x)\gamma_5s(x)$ and $\bar u(x)\gamma_5u(x)$ we could
have used any $\bar c(x)\Gamma s(x)$, $\bar u(x)\Gamma u(x)$ as for any
$\Gamma$ the $s$ channel quantum number are scalar. (In the above
expression 
a $\gamma^\mu\cdot\gamma_\mu$ or $\sigma^{\mu\nu}\sigma_{\mu\nu}$ etc.
contraction is implicit and $\bar c(x)\Gamma s(x)$ 
indicates a contraction over the color indices.)  For
calculations done with infinite precision, it does not matter if we start with
(\ref{e06a}), (\ref{e06b}) or Eq. (\ref{e02}) with $i,j=c,s$,  the lowest energy state in all
cases is the same $D_s(2320)$.

In practice and for motivating the different inequalities there are clear
differences.

In passing we note that a specific member among the three components of
$D_s(2320)$ indicated in Eq. (\ref{e01}) can be enhanced if we probe the physical
$D_s$ at different distance scales.

Thus at short distances the pure $c\bar s$ component may prevail, at
intermediate distances the four-quark single bag state is likely to
dominate and at larger distances yet the two mesons $DK$ 
bound state may be appropriate.

Finally we note that in the case of the $I=1$ analog of our $D_s(2320)$
state say $\bar cs\bar ud$, $\bar cs~{\bar uu-\bar dd\over\sqrt{2}}$,
$\bar cs\bar du$ where all the quarks necessarily propagate from $x$ to
$y$ and only the first ``flavor connected" terms appears in Eq. (\ref{e07a}) and (\ref{e07b}).

For the case of interest with identical $\bar uu$ (or $\bar dd$) annihilation
is possible.  We have the additional second ``flavor disconnected" terms
in (\ref{e07a}), (\ref{e07b}) in which $\bar uu$ annihilate into intermediate gluons and
the quark line emerging from $x$ (pr $y$) loops back to $x$ (or $y$).  The
role of such terms in general and its effect on the inequalities we suggest
in particular will be discussed at length later.  It plays a
crucial role in the pseudoscalar $I=0$ light quark system where 
the axial anomaly coupled to non-perturbative QCD effects \cite{tH} 
explains the absence of a ``ninth" Goldstone boson.

\vskip.5cm

\section {Molecular State via Simple Potential Models.}
\vskip.3cm

Let us assume that $D_s(2320)$ can indeed be viewed as a
$(D^0K^++D^+K^0)/\sqrt{2}$ non-relativistic bound state. This may be
reasonable due to the relatively small binding B.E. $\approx$ 40 
MeV.  There will also be a $D_s\eta^0$ component but being $\simeq$ 200 MeV
heavier will be neglected in sections 3, 4.

We can use a non-relativistic Schr\"odinger equation with a potential generated
by $\rho,\omega$ and the two-meson exchange box diagram (see Fig. \ref{f01}):

\begin{figure}
 \includegraphics{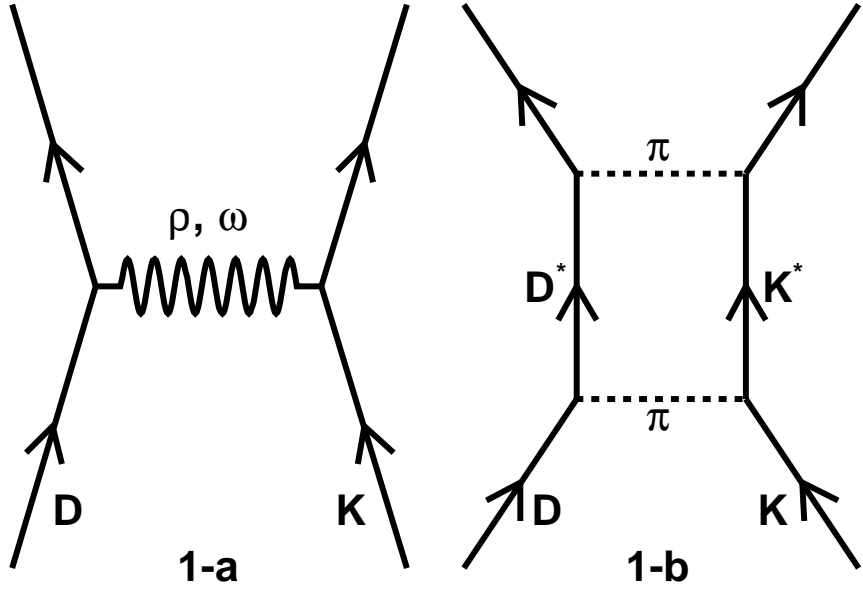}
 \caption{
    \label{f01}
    (a) $\rho / \omega$ exchanges in the $K-D$ channel. (b) 
    The double pion exchange box diagram with intermediate $K^*$ and
    $D^*$
 }
\end{figure}

At the present we do not know $g_{\rho\bar DD}$ $g_{\omega\bar DD}$
and $g_{D^*D\pi}$ to better than $\approx$ 40\%.

It is unrealistic to consider $D$ and $K$ as point-like when the range of
the potential $1/m_v \approx 1/3-1/4$ Fermi, i.e. the distance between the
mesons when $V(r)$ is appreciably, is smaller than the size of each hadron.
Thus we need to include ``Form-Factors" $F_v^D(\vec q)^2$ and
$F_\pi^{DD^*}(\vec q)^2$ and $F_v^K(\vec q)^2,F_\pi^{KK*}(\vec q)^2$
which are poorly known.  Thus any conclusion as the
existence of a $DK$ bound state with $O(40$ MeV) binding is very tentative.

The same applies to the near threshold $\bar KK$ bound states [tentatively
identified with $f_0(980)$ and $a_0(980)$] and even more so to putative
$\bar DD$ $0^+$ bound states.

The main point of the present paper is that certain inequalities between the
masses of the lowest states in the $\bar KK,\bar DD$ and $DK+KD$ channel can
be derived irrespective of all the above ambiguities.

The mass inequality alluded to is:

\begin{equation}
2m_0^{(0^+)}(KD) \geq m_0^{(0^+)}(D\bar D) + m_0^{(0^+)}(\bar KK)
\label{e08}
\end{equation}

We will also attempt to derive directly from the path integral
expressions (\ref{e07a}), (\ref{e07b}), the analog inequalities for quadriquarks:

\begin{equation}
2m_0^{(0^+)}\biggl\{\bar cs{(u\bar u+d\bar d)\over\sqrt{2}}\biggr\} \geq
m_0^{(0^+)}\biggl\{\bar cc{\bar uu+\bar dd\over\sqrt{2}}\biggr\} +
m_0^{(0^+)}\biggl\{\bar ss{\bar uu+d\bar d\over\sqrt{2}}\biggr\}
\label{e09}
\end{equation}

The left-hand side of (\ref{e08}) and (\ref{e09}) is to be identified with the new
$D_s(2320)$ the second mass on the right is taken to be that of
$f^{(0)}(980)$.  Eqs. (\ref{e08}) or (\ref{e09}) then predict the existence of a $0^+$
$D\bar D$ bound state (or $c\bar c{\bar uu+\bar dd\over\sqrt{2}}$ exotic)
at a mass lower than 3660:

\begin{equation}
m^{(0^+)}(\bar DD) \leq 2m~D_s(2320)-mf_0(980) \sim 3660
\label{e10}
\end{equation}

In the limit when both $D$ and $K$ are considered to be a ``heavy-light"
$Q\bar q^\prime$ system the forces between $Q\bar q^\prime$ and 
$\bar Q^\prime q$ due to $\pi,\rho\omega\ldots$ etc. exchanges
coupling to the light degrees of freedom become univiersal and independent
of the heavy quark flavors.

The attractive potentials {\bf guarantee} that
for sufficiently heavy $Q$ and $Q^\prime$ the mesons $Q\bar q^\prime$ and
$\bar Q^\prime q$ will bind. This persists for the genuine
baryonium-like states $Q\bar q^\prime Q^\prime\bar q$.\cite{T,MW,GN} 
The same results are obtained when we view the last problem as binding of
a quadri-quark.  In the $m_Q, m_{Q^\prime}\to\infty$ limit, the $QQ^\prime$
will bind into a $\bar 3$ with Coulombic binding
\begin{equation}
\approx {\alpha_s^2\over 2}{m_Qm_Q^\prime\over m_Q+m_Q^\prime}
\gg\Lambda_{\rm QCD}~.
\label{elqcd}
\end{equation}
The remaining two light anti-quarks will then necessarily form a
combined ``baryon-like" state in which the compact $\{QQ^\prime\}$ behaves
as a yet heavier ``effective $\bar Q$'' with a mass equal to $m_Q+m_{Q^\prime}$.

In the case discussed here with $Q\bar Q^\prime$ (rather than $QQ^\prime$)
heavy quarks the $Q\bar Q^\prime$ will
indeed bind (with twice the binding of $QQ^\prime$) into a compact quarkonium
state which is a color singlet.  The lowest physical state will then be
$(Q\bar Q^\prime) + (q\bar q^\prime)$, i.e. a quarkonium and a 
light quark meson.  
QCD suggests that the residual 
``van der Waals'' multi-gluon exchange interactions between the two color
singlet hadrons is attractive. However due to 
the small size and ``rigidity" of
the quarkonium state the color polarizability of the latter is likely to be
rather small, suppressing the strength of this attraction.  This in
turn makes the actual binding of a light pseudoscalar and the heavy
quarkonium unlikely. 

For the case of $D_s(2320)$, $D+K \approx 2360$ is lighter than $\eta + D_s$.
However, for $D\bar D$ system, $2m_D$ is indeed higher than $m_\eta +
m_{\eta_c}$.  The $D\bar D$ ``state" will thus be unstable with respect to
re-arrangement and decay to $\eta+\eta_c$.

\vskip.5cm

\section {Potential Model Derivations of the Inequality.}
\vskip.3cm

For a first simple motivation of the desired inequality we assume that
the non-relativistic potentials in the $K\bar K~
D\bar K$ and $\bar DD$ channels are the same.\cite{NL,N} 

The lowest eigenvalues of the following three
Hamiltonians:

\begin{equation}
h_{11} =  {\vec p^2\over 2m_1} + {\vec p^2\over 2m_1} +
V(r) 
h_{22} = {\vec p^2\over 2m_2} + {\vec p^2\over 2m_2} + V(r) 
h_{12} = {\vec p^2\over 2m_1} + {\vec p^2\over 2m_2} + V(r) 
\label{e11}
\end{equation}

with $\vec p,\vec r$ the relative moment and distance between the
particles in the respective center mass systems, and $m_1 = m_K,~m_2 = m_D$; are given by the Schrodinger equations:

\begin{equation}
h_{11}|\psi^{(0)}_{1\bar 1}\rangle = \epsilon^{(0)}_{11}
|\psi^{(0)}_{1\bar 1}\rangle 
h_{22}|\psi_{1\bar 2}^{(0)}\rangle = \epsilon^{(0)}_{22}|\psi^{(0)}_{2\bar 2}\rangle 
h_{12}|\psi^0_{1\bar 2}\rangle = \epsilon^{(0)}_{12}|\psi^{(0)}_{1\bar 2}\rangle 
\label{e12}
\end{equation}

$|\psi^{(0)}_{1\bar 1}\rangle,~|\psi^{(0)}_{2\bar
  2}\rangle|\psi^{(0)}_{1\bar 2}\rangle$ being the ground state of the
respective 
Hamiltonian.

Eq. (\ref{e11}) implies the operator relation

\begin{equation}
2h_{12} = h_{11} + h_{12}
\label{e13}
\end{equation}

Taking the expectation value in $\psi^0_{12}$:

\begin{equation}
2\langle\psi^{(0)}_{1\bar 2}|h_{12}|\psi^{(0)}_{1\bar 2}\rangle \equiv
\epsilon^{(0)}_{12} = \langle\psi^{(0)}_{1\bar 2}|h_{11}|\psi^{(0)}_{1\bar 2}
\rangle + \langle\psi^0_{1\bar 2}|h_{22}|\psi^{(0)}_{1\bar 2}\rangle
\label{e14}
\end{equation}

The variational principle implies that the two expectation values on
the right-hand side of Eq. (\ref{e14}) exceed the energies of the respective ground
states.

This then yields

\begin{equation}
2\epsilon^{(0)}_{12} \geq \epsilon^{(0)}_{11} + \epsilon^{(0)}_{22}
\label{e15}
\end{equation}

After adding the rest masses $(2m_1+2m_2)$ to both sides we finally
obtain the desired mass inequality

\begin{equation}
2m^{(0)}_{12} \geq m^{(0)}_{11} + m^{(0)}_{22}
\label{e16}
\end{equation}

i.e. Eq. (\ref{e08}) above.

The operator relation (\ref{e14}) can be projected on any partial $\ell$-wave
subspace.  Hence the mass inequalities hold for each wave separately.
The inequalities also apply when ground state energies are
replaced by the sum of energies of the ground state and the first $n$
radial excitations for any $n$.\cite{NL}

Considering the $D$ (and the $K$) mesons as $\bar Qq$ mesons is clearly
an (extreme) idealization and hence the universality of potentials 
in the
$D\bar D$, $K\bar K$, and $K\bar D$ channels is a rather crude approximation.  Indeed the
coupling $g_{\bar KK/\rho\omega}$ and $g_{KK^*\pi}$ will all differ from the
corresponding $g_{\bar DD/\rho\omega}$ and $g_{DD^*\pi}$, as will the
``form factors" associated with coupling to the $D$ and $K$ of different
sizes.

Nontheless the same mass inequalities
(eq. (\ref{e08})) {\bf continue} to hold subject to much more general and rather
weak assumptions!

The only requirement is that the interactions in the $D\bar D$,~$K\bar K$ and
$\bar KD$ channels can be viewed (in either momentum or configuration space)
as convolutions or ``scalar products" of a generalized ``vector"
associated with couplings to the $K$ line and another vector associated in the
same way with the propagating $D$ line.  The $\rho/\omega$ exchanges and the
more complex interactions due, say, to the
pion box diagram of Fig. \ref{f01} can be viewed as some generalized ``metric"
used for this ``scalar product".  We only need that this metric
be {\bf positive}, viz., attractive potentials.  This is closely related to positive potentials used by
E. Lieb in \cite{Lie}.

We note that the ``scalar product" form holds also for the full propagation
of the $\bar KD$ system and with the various exchanges iterated any number of times.

More generally the amplitude for propagating a, say, $\bar KD$ system
configuration 
at $t=0$ to the same configuration at some imaginary time $t = iT$ can
be expressed in terms of a path integral.

The asymptotic behavior of this joint say $\bar KD$ propagation $\sim{\rm tr}
\{e^{-TH}\}$ is then dominated by the lightest,  bound or threshold,
state in this channel and is proportional to $e^{m^{(0)}_{\bar KD}\cdot
  T} 
\buildrel T\to\infty \over \sim {\rm tr}\{e^{-TH}\}$.  Likewise the asymptotic
behavior of the $D\bar D$ and $K\bar K$ joint propagations are
$e^{-m^{(0)}_{\bar KK}\cdot t}$ and $e^{-m^0_{\bar DD}\cdot T}$ respectively.

When there are no interactions the $K,D$ independently propagate from the
initial $(t=0)$ to final $(t=T)$ configurations.  When the interaction is
switched on the propagation is modulated by the interaction potential at all
intermediate times: $t_1,t_2,\ldots t_n (= \Delta n) \ldots,t_N = N\Delta\equiv
T$. 
(Eventually the $\Delta\to 0$ $N\to\infty$ limit is taken).  Thus the
joint propagation has again the form of a scalar product of $N$ component
vectors with the potentials $V(\vec r_1(t_i)-\vec r_2(t_i))$ playing the
role of the positive metric.\cite{NL,Lie} 

Hence

\begin{eqnarray}
P\{\bar DD(0\to T)\} & \approx & \bar V_D(T)\star V_D(T) \nonumber \\
P\{K\bar K(0\to T)\} & \approx & \bar V_K(T)\star V_K(T) \nonumber \\
P\{\bar DK(0\to T)\} & \approx & \bar V_D(T)\star V_K(T) 
\label{e17}
\end{eqnarray}

and the Schwarz inequality implying:

\begin{equation}
\{P_{\bar DD}(T)\}\{P_{\bar KK}(T)\} \geq (P_{\bar KD}(T))^2
\label{e18}
\end{equation}

yields in the $T\to\infty$ limit the desired inequality.

\vskip.5cm

\section {Mass Inequalities for Four-Quark States from QCD.}
\vskip.3cm

We indicated in Section II how the QCD spectrum can be obtained from Euclidean
correlators, which in turn can be evaluated via path integrals.  A
fundamental property underlying the probabilistic, Monte Carlo, approach
to generating background gauge configuration--is the positivity of the
measure $d\mu(A)$ in Eq. (\ref{e05}) which can be proven in vectorial theories.

The same feature underlies derivation of (QCD) inequalities between
Euclidean correlators or products of such correlators.  These inequalities
when applied in the $|x-y|\to\infty$ limit (where the asymptotic behavior
(see Eq. (\ref{e03})) is dominated by the lightest hadron or hadronic system in the
relevant channel) then leads to mass inequalities of the form
$m_H > m_{H^\prime}$ or $m_A + m_B \leq 2m_C$ respectively.

The conjugation property

\begin{equation}
\gamma_5 S_A(x,y) \gamma_5 = S_A^+(y,x)
\label{e19}
\end{equation}

allows us to write the pseudoscalar correlator:

\begin{equation}
F^{0^-}_{u\bar d}\equiv \langle 0|\bar u(x) \gamma_5 d(x))^+
(\bar u(y)\gamma_5 d(y)|0\rangle 
= \int d\mu(A)~{\rm tr}[\{\gamma_5S_A^{(u)}(x,y) \gamma_5 S^{(d)}(y,x)]
\label{e20}
\end{equation}

as

\begin{equation}
F^{0^-}_{u\bar d}(x-y) = \int d\mu(A)~{\rm tr}[\{S^{(u)}_{(A)}(x,y)\}^+
S^{(d)}_{(A)}(x,y)]
\label{e21}
\end{equation}

In the limit of exact isospin symmetry $m^{(0)}_u = m^{(0)}_d$ and
$\alpha_{em} = 0$, the propagators of the $u$ and $d$ quarks in any background
configuration of gauge fields $A_\mu^{(a)}(x)$ are the same and the
integral becomes a sum of squares: 

\begin{equation}
F^{0^-}_{u\bar d}(x-y) = \int d\mu(A) ~{\rm tr}\{(S_A^q(x,y))^+
S^q_A(x,y)\}
\label{e23}
\end{equation}

Up to overall constants it is larger than two-point correlators of
(light) quark bilinears \cite{Wei} involving $\Gamma$ matrices other than 
$\gamma_5$.  Using the asymptotic behavior we infer that the
pion is the lightest meson:

\begin{equation}
m_\pi \leq m_\rho,~m_{A_1},~~m_\sigma,~m_{f^0},\ldots
\label{e24}
\end{equation}

Next we compare pseudoscalar propagators involving mixed flavors, \cite{Wit83} e.g.

\begin{equation}
\langle 0|(\bar b(x)\gamma_5 c(x)^+(\bar b(y)\gamma_5 c(y))|0\rangle \equiv
F^{(0^-)}_{b\bar c}(x-y)
\label{e25}
\end{equation}

with the corresponding flavor diagonal correlated
$F^{(0^-)}_{b\bar b}(x-y)$ and $F^{(0^-)}_{c\bar c}(x-y)$.  Using the above
path integral representation we have:

\begin{eqnarray}
F^{0^-}_{b\bar b} &=& \int d\mu(A) S^{+(A)}_b(x,y) S_b^A(x,y) \nonumber \\
F^{0^-}_{c\bar c} &=& \int d\mu(A) S^{+(A)}_c(x,y) S_c^{(A)}(x,y) \nonumber \\
F^{0^-}_{b\bar c} &=& \int d\mu(A) S^{+(A)}_b(x,y) S_c^{(A)}(x,y) 
\label{e26}
\end{eqnarray}

The Schwarz inequality then yields

\begin{equation}
F_{b\bar b}^{(0^-)}(x-y)\cdot F^{(0^-)}_{c\bar c}(x-y) \geq
|F^{(0^-)}_{b\bar c}(x-y)|^2
\label{e27}
\end{equation}

and using the asymptotic behavior of the three $F$ functions as
$|x-y|\to\infty$ the interflavor mass inequalities:

\begin{equation}
2m^{(0^-)}_{b\bar c} \geq m^{0^-}_{c\bar c} + m^{0^-}_{b\bar b} \equiv
m_{\eta_c} + m_{\eta_b}
\label{e28}
\end{equation}

In principle such inequalities might be expected for any quark flavors

\begin{equation}
2m^{0^-}(i,j) \geq m^{0^-}_{i\bar i} + m^{0^-}_{j\bar j}
\label{e29}
\end{equation}

These inequalities are reminiscent of the meson meson N.R. bound state mass
inequalities discussed in the previous section.  For the case of heavy quark
flavors such as $i=b,j=c$ where a non-relativistic $Q_i,Q_{\bar j}$
bound state picture may be appropriate, there is indeed a very suggestive
connection:  All QCD interactions in the ${}^1S$ pseudoscalar channel
including the hyperfine chromomagnetic interactions are purely
attractive.  Such potentials correspond to the positive definite interaction
kernel \cite{Lie} used above.

Unfortunately when light flavors,  $u,d$ $I=0$ states are involved, the
above arguments are flawed. The current $\bar q_i(x)\Gamma q_i(x)$ creates from the vacuum a
quark and an identical (anti)-quark. After the fermionic (path)
integration  we have 
not only the flavor connected parts

\begin{equation}
\int d\mu(A)~{\rm tr}\{\Gamma S_A^i(y,x)\} ~{\rm tr}\{\Gamma
S_A^i(y,x)\}
\label{e30}
\end{equation}

considered above but also the flavor disconnected part:

\begin{equation}
\int d\mu(A) ~{\rm tr}(\Gamma S^i_A(xx))\cdot{\rm tr}(\Gamma S_A^i(yy))
\label{e31}
\end{equation}

Each of these terms in (\ref{e31}) traces over single propagators starting at
$x$ (or $y$) and ``looping back" to the initial point.

These contributions seem to be related to $\bar qq$
annihilation into multigluon {``glueball"} color singlet states.  The notion
that such ``hairpin" annihilatiion diagrams can be neglected, namely the
``Zweig rule", dates back to the invention of the quark
model.

This neglect is certainly justified by asymptotic freedom in the case of
heavy quarks and by the fact that many meson nonets are
``ideal" with small mixing between the $\bar ss$ and
${\bar uu+\bar dd\over \sqrt{2}}$ that such annihilations would
generate.

We will next proceed to deriving QCD mass inequalities for four-quark 
correlators adopting first the
``Zweig rule" i.e. by neglecting flavor disconnected contributions,
namely the ${\rm tr}\{D_u^{(A)}(xx)\Gamma\}\cdot{\rm tr}\{D_u^A(yy)\Gamma\}$
term in (\ref{e07a}) and in the analogous expressions for $F_{c\bar cu\bar u}(x,y)$
and $F_{s\bar su\bar u}(x,y)$.  

We use the correlators of products of quark bilinear appearing in (\ref{e06b}) instead of (\ref{e06a}) which is more
appropriate if we view $D_s(2320)$ as a $DK$ bound state since
$c(x)\gamma_5\bar u(x) u(x)\gamma_5\bar s(x)$ creates, when acting on the
vacuum ``localized" $D^0$ and $K^+$ states.

The relevant Schwarz type inequality here is:

\begin{eqnarray}
\int d\mu(A) ~{\rm tr}(S_{s_{(A)}}^+(x,0) S_{u_{(A)}}(x,0))
{\rm tr}(S_{u_{(A)}}^+(x,0)S_{s_{(A)}}(x,0)) 
&\cdot & \nonumber \\
\int d\mu(A)~{\rm tr}(S^+_{c(A)}(x,0)S_{u(A)}(x,0))~
{\rm tr}(S^+_{u(A)}(x,0)S_{c(A)}(x,0)) 
&\geq & \nonumber \\
\left|\int d\mu(A)~{\rm tr}(S_{c(A)}^+(x,0)
S_{u(A)}(x,0))~{\rm tr}(S^+_{u(A)}(x,0)S_{s(A)}(x,0)\right|^2 &{}& 
\label{e32}
\end{eqnarray}

The bound on $m(D_s^0(2320))$ in terms of $m(f_0(980))$ and the mass of a
new $(c\bar u)(\bar cu)$, $I=0$ state would then ensue.

There are no obvious flavor disconnected contributions when we start from
(\ref{e06b}) or from correlators of any of the following scalar products of quark
bilinears: $(\bar u(x)\Gamma Q(x))\cdot(\bar Q^\prime(x)\Gamma u(x))$.
However such terms and in particular $\bar u(x)\gamma_5 u(x)$, yielding
the ${\rm tr}(D^A_u(x,x)\gamma_5)\cdot{\rm tr}(D_u^A(yy)\gamma_5)$
``disconnected" terms in the path integral, do arise if we ``Fierz transform"
any of the above $\Gamma_\alpha\cdot\Gamma^\alpha$.  This reflect
the fact that the channels $D_s\eta^0$ and $\eta_c\eta_0$ do couple to 
the $D_s(2320)~0^+$ state and to the new $c\bar c(\bar uu+\bar dd)~0^+$ state
even if the latter are mainly $KD$ (or $D\bar D$) bound states.

Such disconnected contributions can be avoided if we consider instead the
$I=1$ $(c\bar s~u\bar d)$ $(c\bar c~u\bar d)$ and $(s\bar s~u\bar d~0^+)$
states with the last being the $a_0(980)$ $K\bar K$ threshold state.

However these states can decay into final states involving
the very light pion viz., $D_s\pi^+~\eta_c\pi^+$ and the observed
$\eta\pi$ respectively, with appreciable $Q$ values. This will make
them 
relatively wide and hard to identify. (The fact that $a_0(980)$ still has a
relatively small $\Gamma\approx 50$ MeV width is somewhat encouraging.)

Coming back to the $I=0$ states of interest we need to address the issue of having
flavor disconnected terms in general and those involving $\Gamma = \gamma_5$ in particular.

\ve

\section {Flavor Disconnected Contributions in Pseudoscalar Channels.}
\vskip.3cm

The following section is largely independent of the rest of this paper.
We present it here to preface the discussion of disconnected
contributions to the inequalities between correlators of various
products of quark bilinears. 

Let us compare the Euclidean pseudoscalar correlators with light quarks in
the $I=1$ and $I=0$ channel, used to compute the $\pi,\eta$ masses, respectively

\begin{equation}
\int d\mu(A)~ {\rm tr}\{S_q^{+(A)}(x,y) S_q^{(A)}(x,y)\}
\equiv F^{``\pi"}(x-y)\buildrel|x-y|\to\infty\over\sim
e^{-m_\pi|x-y|} 
\label{e33a}
\end{equation}

\begin{eqnarray}
\int d\mu(A)\bigl[ &{\rm tr}& \{(S_q^{+A}(x,y)S_q^A(x,y))\} + {\rm tr}
(S_q^A(x,x)\gamma_5)\cdot{\rm tr}(S_q^A(y,y)\gamma_5)\bigr] 
\nonumber \\
                   &\equiv  &
F^{``\eta"}(x-y) \buildrel|x-y|\to\infty\over\sim
e^{-m_\eta|x-y|} 
\label{e33b}
\end{eqnarray}

For simplicity we assume only $SU(2)$ flavor with light and
equal mass up and down quarks:

\begin{equation}
m_u^{(0)} = m_d^{(0)} = m_q
\label{e34}
\end{equation}

and further take $\alpha_{\rm em} = 0$.  In this limit $I$-spin is
exact and the $\eta$ particle--all decays of which are electromagnetic and/or
$I$ spin violating--becomes stable. We know experimentally and from
QCD or soft pion / current algebra that in the limit $m_q^0\to 0$, $m_\pi\to 0$ like:

\begin{equation}
m_\pi^2f_\pi^2 \simeq m_q^0\langle \bar qq\rangle \quad
{\rm or} \quad m_\pi^2 \simeq m_q^0~``\Lambda_{\rm QCD}"
\label{e35}
\end{equation}

On the other hand $\eta$ in $SU(2)$ or $\eta^\prime$ in the real world {\bf is}
massive.  The mechanism of generating its mass involves the $U(1)_A$ QCD
anomaly and potentially non-perturbative (instanton) 
effects.  The important point for the present discussion is that
$m_\eta$ (or $m_{\eta^\prime}$) unlike $m_\pi$ does not depend on $m_q^0$ and
stays finite when $m_q^0\to 0$:

\begin{equation}
m_\eta \approx ``\Lambda_{\rm QCD}"
\label{e22}
\end{equation}

and the ratio

\begin{equation}
m_\eta/m_\pi \approx \sqrt{{``\Lambda_{\rm QCD}"\over m_q}}
\label{e36}
\end{equation}

becomes arbitrarily large.

Hence in Eqs. \ref{e33a}, \ref{e33b} above the contribution of the flavor disconnected 
${\rm tr}(S_A\gamma_5)~{\rm tr}(S_A\gamma_5)$ should almost completely cancel
the first positive definite ${\rm tr}(S^+S)$ over a large range of
distance scales:

\begin{equation}
{1\over\Lambda_{\rm QCD}} = m_{\eta^\prime}^{-1} \leq (x-y) \leq
m_\pi^{-1} \approx {1\over (\Lambda_{\rm QCD}m_q^0)^{1/2}}
\label{e37}
\end{equation}

\begin{figure}
 \includegraphics{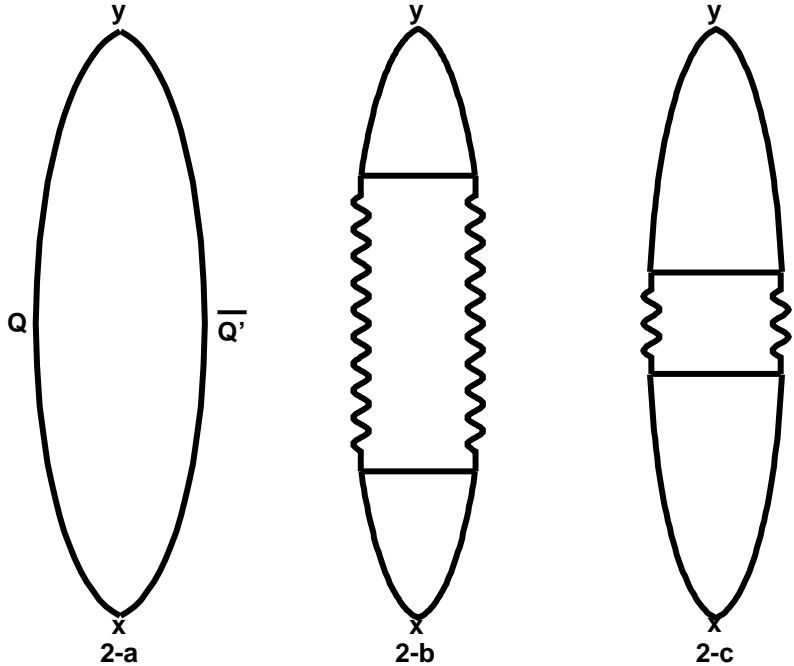}
 \caption{
   \label{f02}
   (a) The propagators in the background gauge field appearing in the
   first, flavor-connected, part of the path integral. (b) The second,
   flavor disconnected part in the path integral. (c) The flavor
   disconnected part redrawn so as to bring out the fact that quark lines
   rather than pure glue propagate most of the way between the points $x$
   and $y$ when $|x-y| >> {\Lambda_{QCD}^{-1}}$.
 }
\end{figure}

If we schematically represent the connected and disconnected parts
in Fig. \ref{f02} 
such a cancellation seems implausible.  Clearly $S_q^A(x,y)$ which appear
in the first term ``knows" via

\begin{equation}
S_q^A(x,y) \equiv \biggl\langle x\biggl|{1\over iD\!\!\!/_A+m_q^0}\biggr|y
\biggr\rangle
\label{e38}
\end{equation}

of both $x$ and $y$ and $m_q^0$.  On the other hand, $S_q^A(x,x)$ and
$S_q^A(y,y)$ ``know" each only $x$ or $y$.  In an ordinary Feynman
diagram sense, there are gluon lines connecting the two loops.  The
propagation of the latter is independent of the light quark masses.
Rather the pure gluonic states propagating between two small loops around $x$
and $y$ would tend to make the contribution of the second flavor
disconnected term $\sim e^{m^{(0^-)}_{GB}|x-y|}$.  Since the lightest
$0^-$ glueball is massive, $m^{(0^-)}_{GB} \sim 2$ GeV $\gg m_\pi$, this second
term could not cancel the first positive term $\sim e^{-m_\pi|x-y|}$
over the interval in Eq. (\ref{e37}) above.  

The resolution of this paradox is very simple.  The pure gluonic state
hardly propagates over any 
distancee.  Rather due to the lightness of the quarks the dominant fermionic
paths making most of the contribution to the propagation and path integral
in (\ref{e07b}) are very extended and go most of the way from $x$ to $y$
(when $|x-y| \gg \Lambda_{\rm QCD}^{-1}$) as indicated in Fig. \ref{f02}c.  Only a
tiny gap remains to be bridged over by the gluons.  

Paranthetically we note that the actual pattern of fermionic paths 
may be 
different from what is suggested by Fig. \ref{f02}.  Thus in the strong
coupling limit large E, B fields are excited and there are in general
few ``channels" 
along which the fermions can propagate with minimal hindrance.  The
absolute square appearing in 
$F_\pi(x-y)$) ($tr~S^+(x,y)S(x,y)$) makes it favorable to have also the
returning anti-quark line follow the same channel.  This 
then would suggest that the worldlines of the quark and antiquark stay
at a finite distance (the width of the above ``channels'' $\sim 
\Lambda_{\rm QCD}^{-1}$) from each other and are ``confined''. It is
amusing to go further \cite{C,BC} and gain a qualitative understanding of SXSB
in this limit.

The vacuum ``channels'' exist independently of the probing at $x$ and
$y$ and the $q_i\bar q_j$ propagation from $x$ to $y$.  These channels
random walk with a finite $(\Lambda_{\rm QCD})^{-1}$ step and often bend
``backwards''. Such bends appear at a given time slice as $\bar qq$
pairs admixed inside the propagating pion.  
This then suggests that a simple
explanation of Eq. (\ref{e35}) follows from a more careful consideration of these $\bar qq$ paths in the
pion.  Similar paths with one small break appear in the case of the $\eta$
in the disconnected parts.  This may naturally cancel the long-range
$e^{-m_\pi|x-y|}$ contribution.

\section {QCD Mass Inequalities for Four-Quark States.}
\vskip.3cm

The disconnected ``hairpin'' diagrams play a crucial role for the $I=0,~0^-$
light pseudoscalar channels and drastically decrease the binding
therein. As we will next argue the effect of such light quark
annihilations on the 
quadriquark mass inequalities may enhance the bindings in the
$Q\bar u\bar Qu$ channels.  Hence the contribution of the second term
in (\ref{e07b}) is likely to be positive and the derivation of the inequalities goes through.

\begin{figure}
 \includegraphics{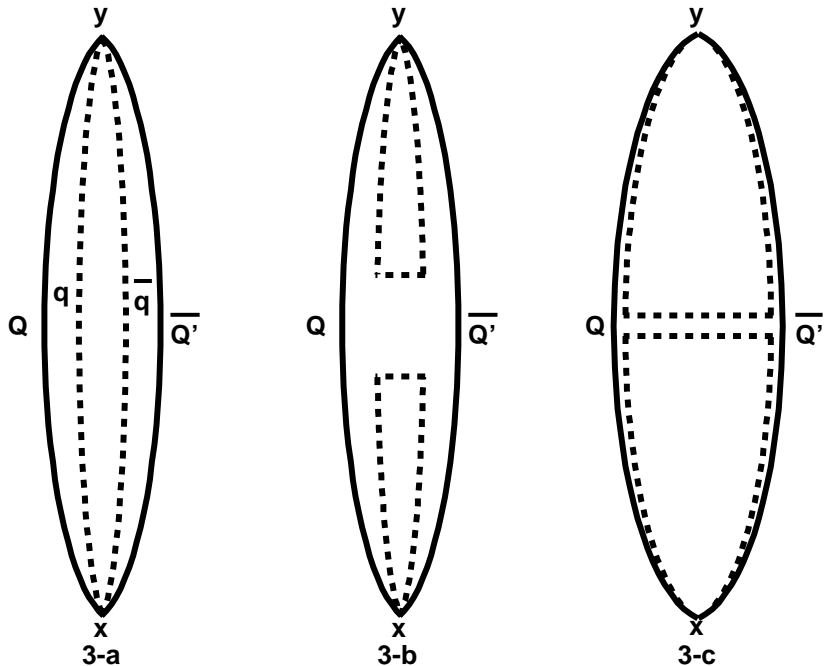}
 \caption{
   \label{f03}
   (a) The first, no-annihilation term.
   (b) The second term with the light $q-\bar q$ quarks 
   annihilating. 
   (c) The second term redrawn to emphasize the short length over 
   which pure glue propagates instead of light quarks, and the connection to
   Harari-Rosner duality diagrams with $\rho$, $\omega$ exchanges.
 }
\end{figure}

To illustrate this point we redraw in Fig. \ref{f03}b the flavor disconnected
terms i.e., the second part in (\ref{e07b}) in a manner similar to that used in
Fig. \ref{f02}a.  Specifically, we let the light quarks propagate over most of
the Euclidean distance $|\vec x-\vec y|$ with a short gap bridged by
gluons.  This would be justified if the four-quark state $Q\bar Q^\prime
q\bar q$ indeed is lighter than the state of $\bar Q Q^\prime$ plus glue
propagating in the middle part of Fig. \ref{f03}b. 
[Note that the lightest such glueball state is a $0^{++}$]

The diagram depicted in Fig. \ref{f03}c is just the classical Harari-Rosner ``Duality
diagram'' with a relatively {\bf light} $\bar qq$ bound state exchanged in the
$t$-channel of the meson-meson scattering.
(The fact that the four-quark lines eventually merge at
$x$ (and at $y$) implies that we project out only the $J^P = 0^+$ state of
interest. It does not affect the following argument pertaining to the
sign of this contribution.)  The lightness is reflected via the narrow gap
between the quark and anti-quark propagating in the $t$-channel.  In a
string model description this separation is proportional to the mass of
the $\bar qq$ meson.  These 
$\omega,\rho,\sigma$ etc. exchanges were indeed considered in our $D\bar K$,
$K\bar K$, and $D\bar D$ potential model picture of Section III, and are
well known to provide {\bf attractive} forces (particularly in the
$\bar MM$ diagonal channels).  Indeed this underlies the argument for positive
interaction kernels used above in the first, heuristic derivation of
the desired mass inequality. For QCD inequalities between four-quark
states, we need positive ``disconnected'' (second)
part in (7b) just like the ``connected" first
part.  We cannot demonstrate this separately for each background field
in the QCD path integral \ref{e07b}. The following arguments are, however,
suggestive. 

First, we compare the joint propagation of four quarks from $x$ to $y$ with
the separate independent propagation of the two $\bar qq$ pairs.  We
argue that with annihilation neglected 

\begin{eqnarray}
&\omit& \langle 0| \{\bar c(x)\gamma_5 u(x)\}\cdot\{\bar c(x)\gamma_5 u(x)\}^+
\cdot\{\bar c(y)\gamma_5 u(y)\}\cdot\{\bar c(y)\gamma_5u(y)\}|0\rangle 
\nonumber \\
&\equiv & F_{D\bar D}(x,y) 
\geq |\langle 0|\{\bar c(x)\gamma_5 u(x)\}^+\cdot\{\bar c(y)\gamma_5
u(s)\}|0\rangle|^2 
\equiv (F_D(x,y))^2 \\
&\omit& \label{e39a}
\end{eqnarray}

and also motivate

\begin{eqnarray}
&\omit& \langle 0| \{\bar c(x)\gamma_5 c(x)\bar u(x)\gamma_5 u(x)\}^+
\cdot(\bar c(y)\gamma_5 c(y))\cdot\bar u(y)\gamma_5 u(y)|0\rangle 
\nonumber \\
 &\equiv & F_{\eta_c\eta}(x,y) 
\geq\langle 0|\bar c(x)\gamma_5 c(x)^+\rangle\bar c\gamma_5(y)|0\rangle
\cdot\langle 0|\{\bar u(x)\gamma_5 u(y)\}^+\bar u(y)\gamma_5 u(y)
\nonumber \\
&=& F_{\eta_c}(x,y)\cdot F_\eta(x,y) \\
&\omit& \label{e39b}
\end{eqnarray}

In the $|x-y|\to\infty$ limit
these may suggest that bound $c\bar u-\bar cu$ or
$c\bar c-u\bar u$ states exist.

\begin{equation}
m^{0^+}(c\bar u,\bar cu) \leq 2m^{0^+}(c\bar u) 
\label{e40a} 
\end{equation}

\begin{equation}
m^{0^+}(c\bar c,u\bar u) \leq m^{0^+}(c\bar c) + m^{0^+} u\bar u)
\label{e40b}
\end{equation}

With disconnected terms omitted:
\begin{equation}
F_{D\bar D}(x,y) \equiv \int d\mu(A) ~{\rm tr}\{S_c^{+(A)}(x,y))~S_u^{(A)}(x,y))\}
{\rm tr}\{S_u^{+(A)}(x,y) S_c^{(A)}(x,y))\}
\label{e40c}
\end{equation}

\begin{equation}
F_D(x,y) = \int d\mu(A)~{\rm tr}~S_c^{+(A)}(x,y) S_u^{(A)}(x,y)
\label{e41a}
\end{equation}

and

\begin{eqnarray}
F_{\eta_c\eta}(x,y) &=& \int d\mu(A)~{\rm tr}(S_c^{+(A)}(x,y)~{\rm tr}
\{S_u^{+(A)}(x,y) S_u^{(A)}(x,y)\} \nonumber \\
F_{\eta_c}(x,y) &=& \int d\mu(A)~{\rm tr}\{~S_c^{+(A)}(x,y)~S_c^{+(A)}(x,y)\}
S_c^{(A)}(x,y)\\
F_\eta(x,y) &=& {\rm tr}(S_u^+(x,y) S_u(x,y)
\label{e41b}
\end{eqnarray}

The inequalities

\begin{equation}
F_{\bar DD}(x,y) \geq |F_0(x,y)|^2 
\label{e42a} 
\end{equation}

and

\begin{equation}
F_{\eta_c\eta}(x,y) \geq F_{\eta_c}(x,y) F_\eta(x,y) 
\label{e42b}
\end{equation}

then follow either as a Schwarz inequality or from more heuristic
considerations in the case of equation \ref{e42b} (see footnote).  

These correlator 
inequalities imply the mass inequalities (\ref{e40a}, \ref{e40b}).  A more careful
analysis \cite{NL,NS} suggests that the latter {\bf do not} imply that 
$D\bar D$ and/or $\eta_c\eta$ bound states exist.  Rather it implies an
attractive $S$-wave scattering length.\footnote[1]{
The second inequality holds if $|S_c^i(x,y)|^2$ and $|S_u^i(x,y)||^2$
are positively correlated; namely for those $i$ values; i.e. $A_\mu^a(x)$
background field configurations, where $|S_c^i(x,y)|^2$ is larger/smaller
than the average also $|S_u^i(x,y)|$ tends to be large/small.  Indeed those
configurations which do (not) impede light quark ($u$) propagation are
likely not to impede that of the heavy quark $(c)$.  The heuristic statement
becomes much clearer when we have quarks of the same mass and indeed we
expect the Casimir polder type attraction to be maximal between objects with
similar ratios $\alpha_M/\alpha_E$ of the magnetic and electric
polarizability.}

What is the impact of turning on the flavor disconnected contributions
on the 
mass inequalities?  One 
effect is the enhanced $\eta$ mass (relative to $\pi^0$).  This however
leaves attractive
interactions between hadrons in general and between the $\eta$ and $D_s$ and $\eta$
and $\eta_c$ in particular. The interaction in the last
case involves only gluonic exchanges as there are 
no common light quark to annihilate and/or exchange.  Such two (or more)
gluon exchange ``Casimir-Polder" type interactions between objects of similar
chromo-electric magnetic properties (as {\bf any} two hadrons invariably are)
tend to be always attractive.

The transitions to states including $\eta,\eta'$ appears to be
the main obstruction to both the derivations of the inequalities (which
follow if we consider only the $\bar DD$, $\bar KK$, and $\bar KD$
channels or the corresponding four quark correlators).  The very same
transition may impede the discovery of the predicted new
$\bar DD$ bound state as the latter may have a broad decay width
into $\eta_c+\eta$.  Indeed heavier $Q,Q'$ quarks in the $Q\bar Qq\bar q$ state
considered make such a transition more likely.  Thus the $D_s(2320)$
bound by $\simeq 40$ MeV in the $D\bar K$ channel the $D_s\eta$
threshold is yet $\approx 200$ MeV higher.  However the new $D\bar D$
state which our inequality predicts is $\simeq 100$ MeV below the $\bar
DD$ threshold and  $80 MeV$ above the $\eta_c\eta$ threshold.   

The relatively narrow width ($\sim$ 50 MeV) of $f_0(980)$ into $\pi\pi$
states (where annihilation of slightly  heavier $\bar ss$ is involved and the similar width of
$a_0(980)$ into $\eta\pi^0$ (where no $\bar ss$ annihilation are needed)
suggests that also the decay width of the new $\bar DD(2660)$ ${0^+}$
state into $\eta_c\eta$ may not be 
prohibitively large.  

The above discussion suggests attractive interactions in both $\bar DD$
and $\eta_c\eta$ 
channels.  Hence even if the $\eta_c\eta$ channel does not bind on its
own it generates some threshold enhancement.  The mixing 
of the $\eta_c\eta$ and $\bar DD$ channel may in turn further lower the
mass of the state of interest.

All the above considerations then suggest that the new $\bar DD$
state predicted by our QCD inequalities -- while definitely broader than
the narrow $D_s(2320)$ state may still manifest in high statistcs
experiments.  Before 
concluding we add few more comments:

(i) Obviously $B_s$ states analogous to the $D_s(2320)$ exist. 
The $\bar BB$ analog of our predicted $\bar DD$ state 
lies even higher above the $\eta\eta_b$ threshold. The 
small chromo-electric polarizability of $\eta_b$ will strongly quench the
``Casimir-Polder" residual interactions between $\eta_b$ and $\eta$.

(ii) An issue omitted in the above in the role played by annihilation of
 the ``heavy'' $\bar ss$ pairs in $f_0(980)$ which again is assumed to be an
 $\bar ss(\bar uu+ \bar dd)$ $0^{++}$, $I=0$ state. The relative small
 $\Gamma\sim50 MeV$ of $f_0(980)\rightarrow\pi\pi$ $S$-wave decay
 despite the huge Q value indicates that to a large extent the ``Zweig
 rule'' is operative and the annihilation of $\bar ss$ pairs is
 suppressed. 

(iii) It has been suggested that the new BaBar relatively light $D_s(0^+)$
state could be interpreted as the ``parity doublet partner" of
$D_s$ ${(0^-)}$ state.\cite{BEH} 
Parity doubling is a vestige 
of a world where the $S \chi SB$ transition has not occurred and
the $Q_5$ symmetry generated via $I_5^\pm = \int d^3x \bar u(x)\gamma_5\gamma_0
d(x)$; or $\bar d(x)\gamma_5\gamma_0u(x)$ and $I_5^3$ is linearly realized
via parity doublets.  Superficially this seems to be an altogether different
interpretation than the states $Q\bar Q^\prime\bar qq$ with one extra pair
of light quarks considered here.  This need not be the case.  Parity doubling, i.e. degeneracy of $S$ and
$P$-wave states is extremely difficult to understand in a simple N.R.
quark model due to the prohibitive cost of $P$-wave $q\bar q$ excitations
(a point which we belabored in the first section).  Indeed unbroken
chiral symmetry does not hold for most composite models.
We can incorporate this feature
by assuming that hadrons contain an ``infinite" number of
$\bar qq$ pairs.  In this case $I_5^{(+)}$ when acting on a fermion, generates 
an extra $(\bar ud)$ pair which blends into the
``coherent" state with many pairs producing the parity doublet
without any energy penalty.

The introduction of a light $\bar qq$ pair into the $c\bar s$ (or $Q^\prime
\bar Q$) $0^-$ ground state was motivated above as a way to achieve a lower
energy realization of a $0^+$ state than via a
simple $P$-wave excitation.  This can be viewed as a modest, first step towards 
the infinitely many light quark anti-quark pairs 
implicit when ideal parity doubling holds.

The spectroscopy of the $c\bar c$ system is well understood
 and ALL the four P-wave states are accounted for. Finding below $D\bar D$
 threshold the new $c\bar cu\bar u$ I=0  0$^{++}$ state that QCD inequalities
 suggest (if the  four quark interpretation of $D_s$(2320) and $f_0$(980)
 is correct) would constitute compelling evidence for four quark states.

At the present time the discovery of one of the missing 1$^+$ $c\bar s$ 
state at 2460 MeV is deemed to favor interpreting  $D_s$(2320) as a P-wave excitation
 of two quarks. Clearly such an interpretation will be bolstered if the
missing heavier and broad 0(+) $c \bar s$ state and the new $\bar cc$
state discussed here will not be discovered. 

We note that the KM favoured $b\to c\bar cs$ constituting $\sim$20\% of all
B decays may offer good hunting grounds for the new particle which
predominantly decays into $\eta_c \eta$. 

\section{Acknowledgements}

I would like to thank M.~Cornwall for helpful
discussions on the $\eta-\pi$ problem and M.~V.~Purohit for insight on the
prospects of experimentally discovering the proposed state. I am
indebted to R. Schrock for many helpful on quark models.

\end{document}